\documentclass{aastex6}
\usepackage{mkfig}

\begin{document}

\title{Inferences About the Magnetic Field Structure of a CME
  with Both In~Situ and Faraday Rotation Constraints} 

\author{Brian E. Wood\altaffilmark{1}, Samuel Tun-Beltran\altaffilmark{1},
  Jason E. Kooi\altaffilmark{2}, Emil J. Polisensky\altaffilmark{2},
  Teresa Nieves-Chinchilla\altaffilmark{3}}
\altaffiltext{1}{Naval Research Laboratory, Space Science Division,
  Washington, DC 20375, USA; brian.wood@nrl.navy.mil}
\altaffiltext{2}{Naval Research Laboratory, Remote Sensing Division,
  Washington, DC 20375, USA}
\altaffiltext{3}{Heliophysics Science Division, NASA Goddard Space
  Flight Center, Greenbelt, MD 20771, USA}


\begin{abstract}

     On 2012~August~2, two CMEs (CME-1 and CME-2) erupted from the
west limb of the Sun as viewed from Earth, and were observed in
images from the white light coronagraphs on the {\em SOHO} and {\em STEREO}
spacecraft.  These events were also observed by the Very Large Array
(VLA), which was monitoring the Sun at radio wavelengths, allowing
time-dependent Faraday rotation observations to be made of both
events.  We use the white-light imaging and radio data to model the
3-D field geometry of both CMEs, assuming a magnetic flux rope
geometry.  For CME-2, we also consider 1~au in~situ field measurements
in the analysis, as this CME hits {\em STEREO-A} on August~6,
making this the first CME with observational constraints from stereoscopic
coronal imaging, radio Faraday rotation, and in~situ plasma
measurements combined.  The imaging and in~situ observations of CME-2
provide two clear predictions for the radio data; namely that VLA
should observe positive rotation measures (RMs) when the radio line of
sight first encounters the CME, and that the sign should reverse to
negative within a couple
hours.  The initial positive RMs are in fact observed.  The expected
sign reversal is not, but the VLA data unfortunately end too soon to
be sure of the significance of this discrepancy.  We interpret an RM
increase prior to the expected occultation time of the CME as a
signature of a sheath region of deflected field ahead of the CME
itself.

\end{abstract}


\section{Introduction}

     Coronal mass ejections (CMEs) are among the most dramatic of
solar eruptive phenomena, which can have significant geomagnetic
effects when directed toward Earth.  Assessing the structure of CMEs
and predicting their geoeffectiveness has therefore been a central
goal of heliophysics.  The primary observational diagnostics of CMEs
are white-light images of the transients.  The most prodigious source
of these observations over the past two decades is the Large Angle
Spectrometric COronagraph (LASCO) instrument on board the {\em SOlar
and Heliospheric Observatory} ({\em SOHO}).  From {\em SOHO}'s vantage
point at the Sun-Earth L1 Lagrangian point, the LASCO C2 and C3
coronagraphs have been monitoring the solar corona within 30~R$_{\odot}$
of the Sun since 1996 \citep[e.g.,][]{sy04,ng09,er09}.

     However, interpreting images of CMEs can be difficult, due to the
problem of inferring three-dimensional structure from two-dimensional
images.  Although CMEs are often described in terms of a three-part
structure, with a bright core near the back end of a dark cavity
surrounded by a bright loop \citep{rmei85,seg98},
their appearance can vary tremendously from event to event.  A
crucial advance in CME studies has been the advent of stereoscopic
imaging of CMEs, particularly using observations from the {\em Solar
TErrestrial RElations Observatory} ({\em STEREO}) mission, with twin
spacecraft ({\em STEREO-A} and {\em STEREO-B}) monitoring the solar
corona and interplanetary space from two separate vantage points 1~au
from the Sun.  Analyses of these data have provided support for the
magnetic flux rope (MFR) paradigm of CME structure, where CMEs are
assumed to be long tubes permeated by a helical magnetic field, with
the legs of the tubes stretching back toward the Sun
\citep{at09,ekjk12,av14,bew17,cm18}.
Stereoscopic analysis of CMEs has
recently been extended to include observations from close to the Sun
by the {\em Parker Solar Probe} \citep{bew20}.

     Stereoscopic imaging may provide excellent constraints on a
CME's 3-D shape and kinematics, if the viewing geometry is
advantageous, but this by itself does not elucidate the magnetic
structure of the event.  Images only show mass within the CME, which
may not relate well to the magnetic field that provides the framework
of the CME.  Furthermore, it is the magnetic field strength and
orientation that largely determines whether an Earth-directed CME will
be geoeffective or not.  Thus, observational constraints on magnetic
fields within CMEs are crucial for studies of CME structure.

     The most ubiquitous source of field measurements of CMEs
are magnetometers that are on board many heliophysics missions,
including {\em STEREO}, {\em Wind}, and {\em Advanced Composition
Explorer} ({\em ACE}); the latter two operating near Earth at L1.
Like the imaging data, in~situ plasma and field measurements also
provide support for the MFR paradigm of CME structure, with many
interplanetary CMEs (i.e., ICMEs) characterized by a region of low
plasma $\beta$ and strong, rotating magnetic field, consistent with
the MFR picture \citep{lb81,km86,lfb88,vb98}.
The spacecraft field measurements
suffer from one obvious drawback, and that is that they only provide
measurements for a single track through the CME structure.  Techniques
have been developed to extrapolate from that track to a full 3-D MFR
structure, but such extrapolations require simplifying assumptions
about the physics and geometry of the MFR structure
\citep[e.g.,][]{rpl90,rpl11,rpl15,qh01,mv03,nah13,tnc16,tnc18}.

     Radio observations potentially provide an alternative source
of information about CME magnetic field structure, specifically by
measuring the Faraday rotation induced by a CME passing in
front of a background radio source.  There is a long history of using
Faraday rotation to study coronal plasma
\citep[e.g.,][]{mkb80,ts94,ldi07,smo07,jek14},
but measuring CMEs is naturally harder due to their
unpredictable and transient nature.  The first serendipitous
detections involved radio signals from {\em Pioneer 9} \citep{gsl69}.
A more systematic effort specifically designed to detect CMEs
was made involving {\em Helios}, resulting in five detections \citep{mkb85}.
More recently, there has been a study utilizing the
signal from the {\em MESSENGER} spacecraft \citep{eaj18}.
Detections using astrophysical background sources are few.
\citet{tah16} failed to detect a rotation measure signal in Very
Large Array (VLA) observations of a pulsar that was occulted by a CME,
although limits were placed on the CME's field strength.
However, \citep{jek17} reported successful VLA detections of three
CMEs on 2012~August~2, and
the Low Frequency Array (LOFAR) detected a rotation measure signal
even farther from the Sun for an event on
2014~August~13 \citep{mmb16}.

     We have recently recognized that one of the events observed by
VLA on 2012~August~2 was directed right at {\em STEREO-A}, and hit the
spacecraft on 2012~August~6, where it was observed by the spacecraft's
particle and field detectors.  The {\em STEREO} viewing geometry of
the event was also excellent, at least while near the Sun, providing
the necessary data for an accurate morphological reconstruction.  All
this makes this event truly unique, as being the only CME with
observational constraints from both stereoscopic imaging and Faraday
rotation observations near the Sun, and also in~situ plasma
measurements at 1~au.  This is therefore the ideal event for exploring
CME field structure, and we here present our study of the CME
considering all of the observational constraints as listed.

     The analysis begins in Section~2 with a description of our
morphological and kinematic measurements based on the stereoscopic
imaging constraints.  The morphological reconstruction is based on the
MFR paradigm for CME structure.  There are actually two distinct CMEs
that erupt within two hours of each other off the west limb of the
Sun, as viewed from Earth.  Although we focus mostly on the second CME
that ultimately hits {\em STEREO-A}, the first CME is analyzed as well, as
radio Faraday rotation constraints are available for both CMEs using
two separate background radio sources.  In Section~3, we present
{\em STEREO-A}'s in~situ plasma and field measurements of the second CME,
taken on 2012~August~6 when that CME hits the spacecraft.  These data
provide the basis by which we insert a physically realistic field
structure into the MFR morphology inferred from the imaging.  This
analysis utilizes the new MFR model of \citet[][hereafter TNC18]{tnc18}.
In Section~4, we focus on the radio Faraday
rotation data, with a primary goal of determining whether the MFR
reconstruction of the CME morphology and field structure from the
imaging and 1~au in~situ data is consistent with the radio
constraints.  Finally, in Section~5 we summarize our results.

\section{Reconstruction of CME Morphology from Stereoscopic Imaging}

\begin{figure}[t]
\plotfiddle{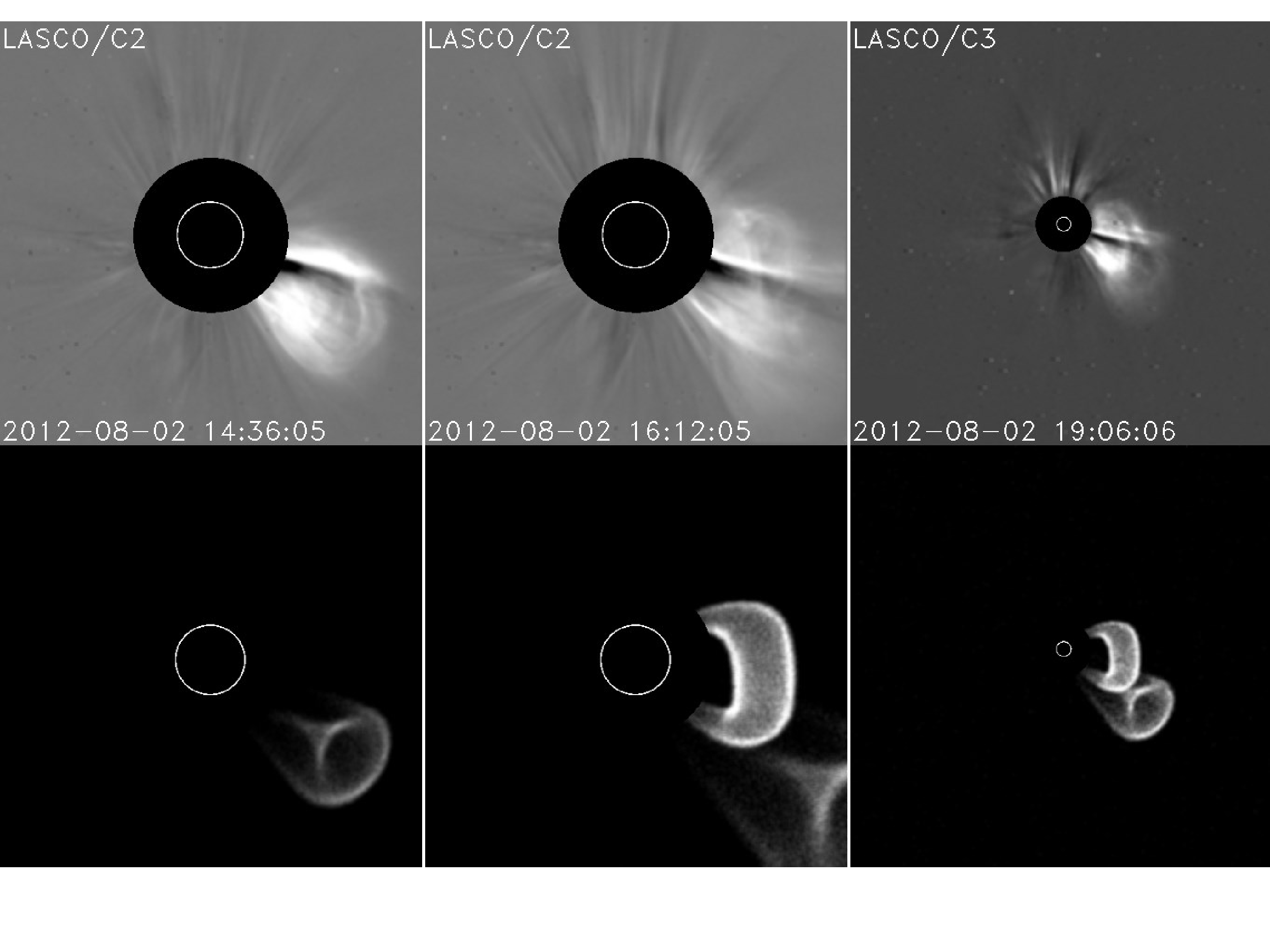}{3.4in}{0}{55}{55}{-200}{-35}
\caption{Images of the 2012~August~2 CMEs from {\em SOHO}/LASCO's C2 and C3
  corongraphs.  The first C2 image shows CME-1, the second CME-2,
  and the third image shows the
  overlapping CMEs in the C3 FOV.  Synthetic images of the event are
  shown below the real images, based on the 3-D reconstruction shown
  in Figure~4.  A movie version of this figure is available online.}
\end{figure}
     The coronagraphic observations of the CMEs on 2012~August~2
were first described by \citet{jek17}.  These include
observations from {\em SOHO}/LASCO's C2 and C3 coronagraphs, covering
plane-of-sky distances from Sun-center of 1.5--6 R$_{\odot}$ and
3.7--30 R$_{\odot}$, respectively \citep{geb95}; and also
by the COR1 and COR2 coronagraphs on board the twin {\em STEREO} spacecraft,
covering distances of 1.4--4.0 R$_{\odot}$ and 2.5--15.6 R$_{\odot}$,
respectively, which are constitutents of {\em STEREO}'s Sun-Earth Connection
Coronal and Heliospheric Investigation (SECCHI) imaging package
\citep{rah08}.  Utilizing the CME identifiers of \citet{jek17},
CME-1 was first seen at UT 13:25 by LASCO/C2,
directed in a roughly southwesterly direction.  This CME is followed
shortly thereafter by CME-2 at UT 14:48, which overlaps CME-1 but with
a trajectory more directly to the west.  This is shown explicitly in
the first two panels of Figure~1, with the third panel showing the
overlapping CMEs in the field of view (FOV) of LASCO's C3 coronagraph.
\citet{jek17} discuss a narrow third CME, CME-3, directed to the
northwest at UT 16:36, which we do not consider in this article.

\begin{figure}[t]
\plotfiddle{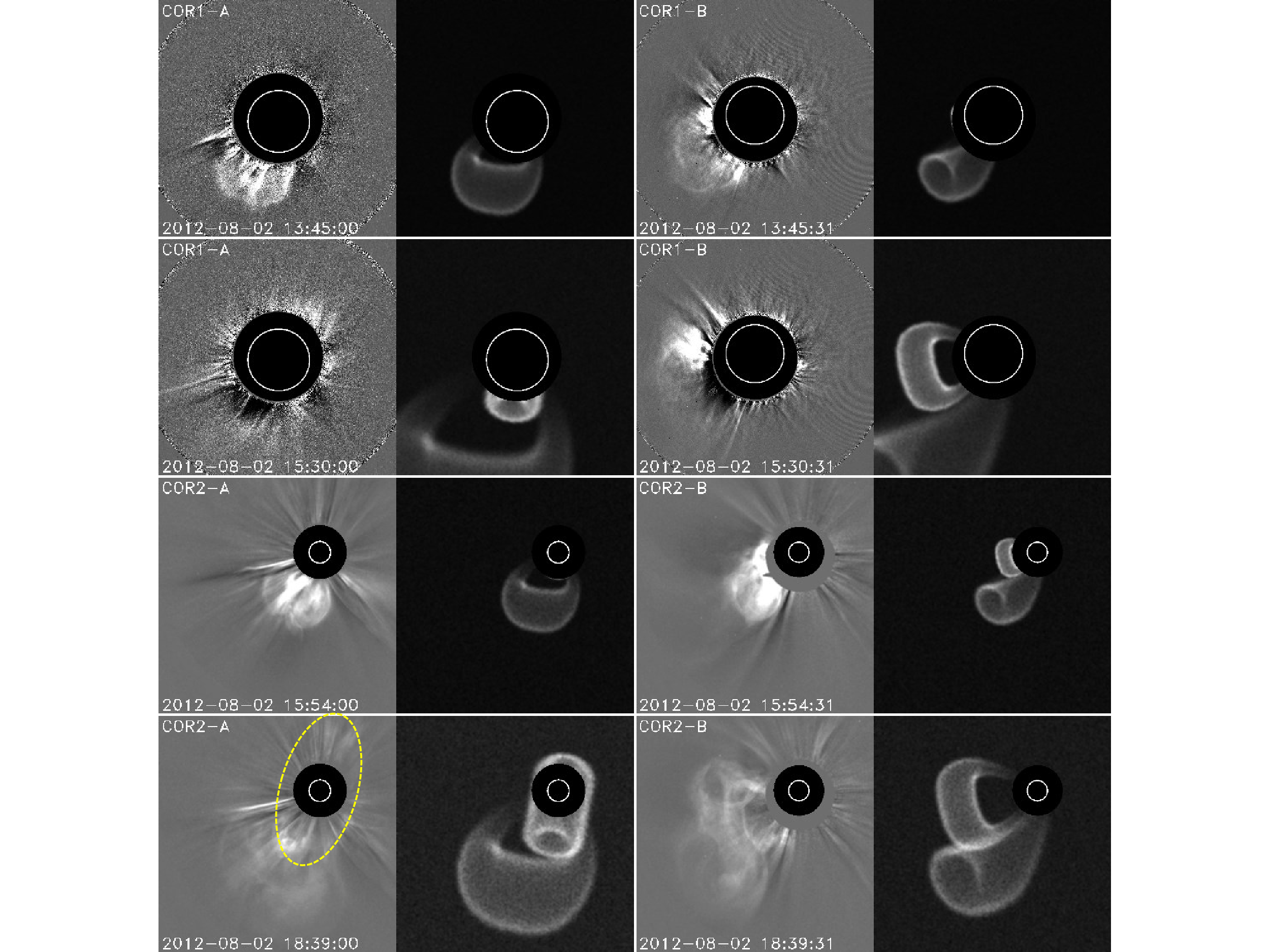}{4.0in}{0}{60}{60}{-215}{-5}
\caption{A sequence of four synchronised images of the 2012~August~2
  CMEs from the COR1 and COR2 coronagraphs on {\em STEREO-A} (left) and
  {\em STEREO-B} (right).  The first COR1 image shows CME-1, and the second
  CME-2, which follows right behind CME-1.  The third image shows
  primarily CME-1 in the COR2 FOV, and the final image shows the
  overlapping CMEs in COR2.  For clarity, a yellow dashed line is used
  in the final COR2-A image to outline CME-2's location.  Synthetic
  images of the event are shown to the right of the real images, based
  on the 3-D reconstruction shown in Figure~4.  A movie version of
  this figure is available online.}  \end{figure}
\begin{figure}[t]
\plotfiddle{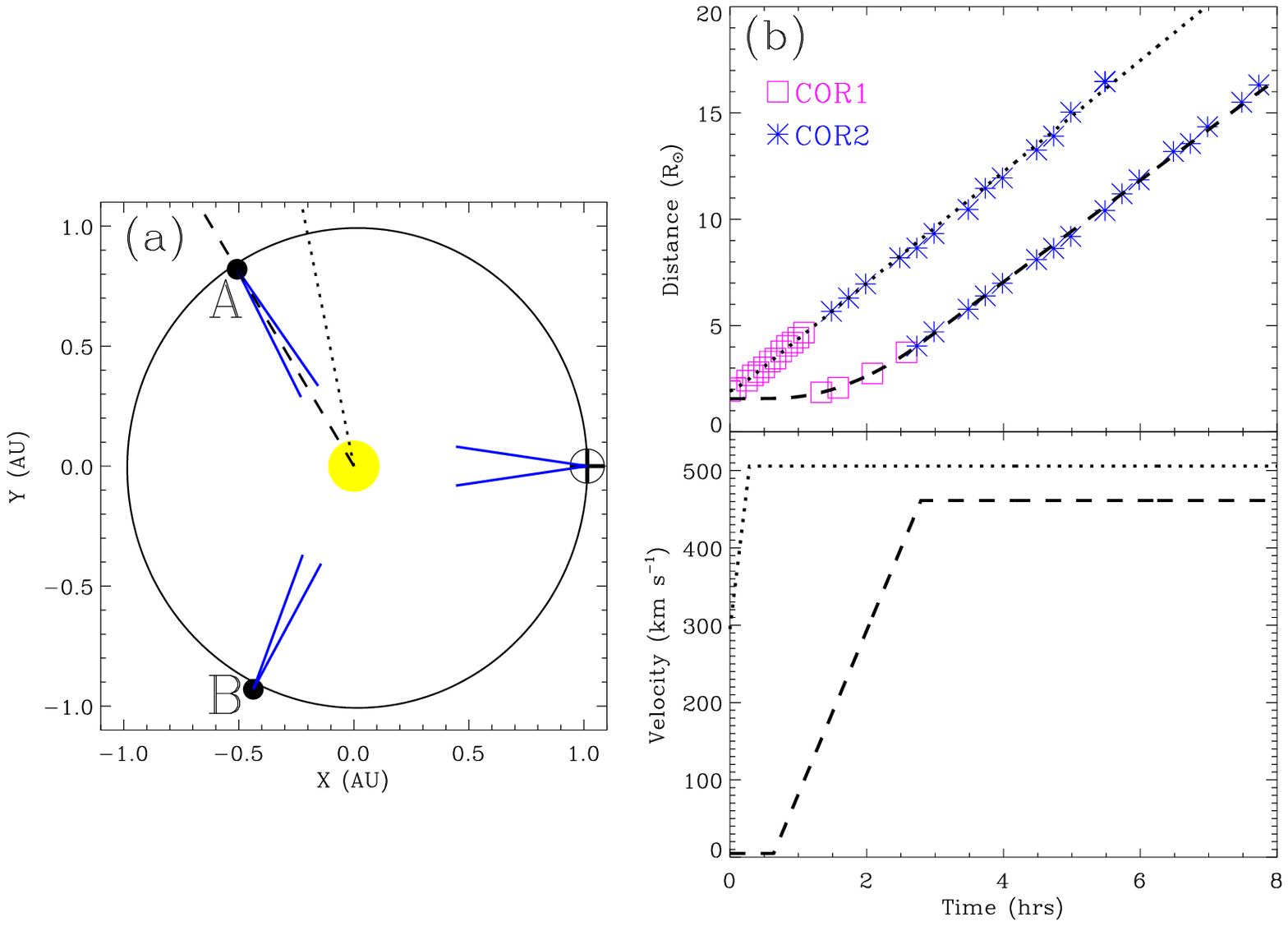}{3.2in}{0}{75}{75}{-240}{-290}
\caption{(a) The positions of Earth, {\em STEREO-A}, and {\em
  STEREO-B} in the ecliptic plane on 2012~August~5 (in HEE
  coordinates).  From each of these perspectives the blue lines show
  the fields of view of {\em SOHO}/LASCO's C3 coronagraph, and the COR2-A and
  COR2-B coronagraphs on {\em STEREO-A} and {\em -B}, respectively.  The dotted
  and dashed lines indicate the central trajectory longitude of CME-1 and
  CME-2, respectively.  (b) The top panel shows distance measurements
  for the leading edges of CME-1 and CME-2 based on images from
  {\em STEREO-B}'s coronagraphs (COR1-B and COR2-B).  The $t=0$ point on the
  time axis corresponds to UT 12:55 on August~2.  These data are
  fitted with a simple kinematic model assuming a constant
  acceleration phase followed by a constant velocity phase.  The dotted
  and dashed lines are the best fits.
  The corresponding dotted and dashed lines in the bottom
  panel show the inferred velocity profiles for CME-1 and CME-2,
  respectively.}
\end{figure}
     Figure~2 shows a sequence of four synchronised COR1 and COR2
images of the two CMEs from the two {\em STEREO} spacecraft.  Interpretation
of the images requires knowledge of the viewing geometry.  This is
shown explicitly in Figure~3(a), which illustrates the positions of
the various observing locations in a heliocentric Earth ecliptic (HEE)
coordinate system, with the x-axis pointed toward Earth and the
z-axis pointed toward ecliptic north.  The COR2-B image at UT 18:39
greatly resembles the LASCO/C3 image at UT 19:06 in Figure~1, but with
the CMEs on the east side of the Sun instead of the west.  This
implies that the CMEs are directed roughly halfway between {\em SOHO}/LASCO
and {\em STEREO-B}, which places their trajectories roughly toward the
longitude of {\em STEREO-A} (see Figure~3a).  The {\em STEREO-A} images provide
support for this conclusion.  From {\em STEREO-A}'s perspective, CME-1 seems
directed close to due south, but slightly to the east.  Finding CME-2
in the {\em STEREO-A} images is trickier.  In the COR2-A image at UT 18:39
in Figure~2, CME-2 is apparent as a faint elliptical halo CME
surrounding the Sun.  This is more apparent in the movie version of
the figure.  (Movie versions of both Figures~1 and 2 are available in
the online version of the article.)  The halo appearance of the CME
indicates that the CME is directed right at {\em STEREO-A}, a
trajectory direction supported by the combined appearance of the CME in
{\em STEREO-B} and LASCO.  Furthermore, CME-2 in fact hits {\em STEREO-A}
four days later on August~6, as will be discussed in Section~3.

     We measure the kinematics of the two CMEs using the {\em STEREO-B}
images, with results shown in Figure~3(b).  We first measure the
elongation angle, $\epsilon$, of each CME's leading edge from
Sun-center as a function of time.  Following past practice
\citep[e.g.,][]{bew10,bew17}, these angles are converted to physical
distances using the prescription of \citet{nl09},
\begin{equation}
r=\frac{2d\sin \epsilon}{1+\sin(\epsilon+\phi)},
\end{equation}
where $d$ is the distance from the observer ({\em STEREO-B} in this case) to
the Sun and $\phi$ is the angle between the CME trajectory and the
observer's line-of-sight (LOS) to the Sun.  This equation assumes the CME can be
approximated as a sphere centered halfway between the Sun and the
CME's leading edge.  The central trajectory and $\phi$ values for the
two CMEs are ultimately inferred from the morphological analysis
described below, with the central longitude of those trajectories
shown explicitly in Figure~3(a).  The CME distances are fitted with a
simple two-phase kinematic model assuming an initial phase of constant
acceleration followed by a coast phase of constant velocity, with the
results shown in Figure~3(b).  Neither CME is very fast, with CME-1
reaching a peak speed of 506 km~s$^{-1}$, and CME-2 reaching 461
km~s$^{-1}$.

     Both {\em STEREO-A} and {\em -B} possess heliospheric imagers that are
often able to track CMEs into interplanetary space, potentially all
the way to 1~au \citep[e.g.,][]{rah08,cje09,bew17}.
However, the trajectories of the two 2012~August~2 CMEs do not
place them in the fields of view of the heliospheric imagers, which are
pointed at the Sun-Earth line.  The central longitude of the CME-1 trajectory
shown in Figure~3(a) suggests that it might be in the heliospheric
imager FOV for {\em STEREO-A}, but in reality the CME trajectory latitude is
too far to the south.  The lack of heliospheric imager constraints
means that our kinematic models of the two CMEs are based only on the
coronagraphic observations close to the Sun.  Nevertheless, in
Section~3 we will show that the kinematic model of CME-2 in
Figure~3(b) assuming that the CME maintains the 461 km~s$^{-1}$ speed
all the way to {\em STEREO-A} predicts the CME arrival time
at {\em STEREO-A} surprisingly well.  This success probably
indicates that the CME speed was close enough to that of the ambient
solar wind to allow the CME to reach 1~au without further acceleration
or deceleration.

     We use the stereoscopic imagery of each CME to reconstruct
its 3-D morphology, asssuming an underlying MFR shape.  For this
purpose, we use a parametrized mathematical prescription for
generating 3-D MFR shapes that we have used many times in the past.
The prescription is described in detail by \citet{bew09},
but is most extensively utilized in the {\em STEREO} CME survey of
\citet{bew17}.  An assumed set of MFR parameters yields a 3-D
MFR shape.  This shape is then used to generate a 3-D density cube,
with mass placed only on the surface of the MFR to outline the CME
boundary, and not in the interior.  We assume self-similar expansion
for the MFR, meaning that this single density cube applies at all
times, with only the axis scale changing in a manner described by the
kinematic model of the CME (e.g., Figure~3b).  Thus, from the 3-D
density cube we can compute synthetic images of the CME for any FOV
from any perspective at any time, for comparison with the actual
images of the CME.  The synthetic images are computed using a white
light rendering routine that precisely computes the Thomson scattering
within the density cube \citep{deb66,afrt06}.

\begin{figure}[t]
\plotfiddle{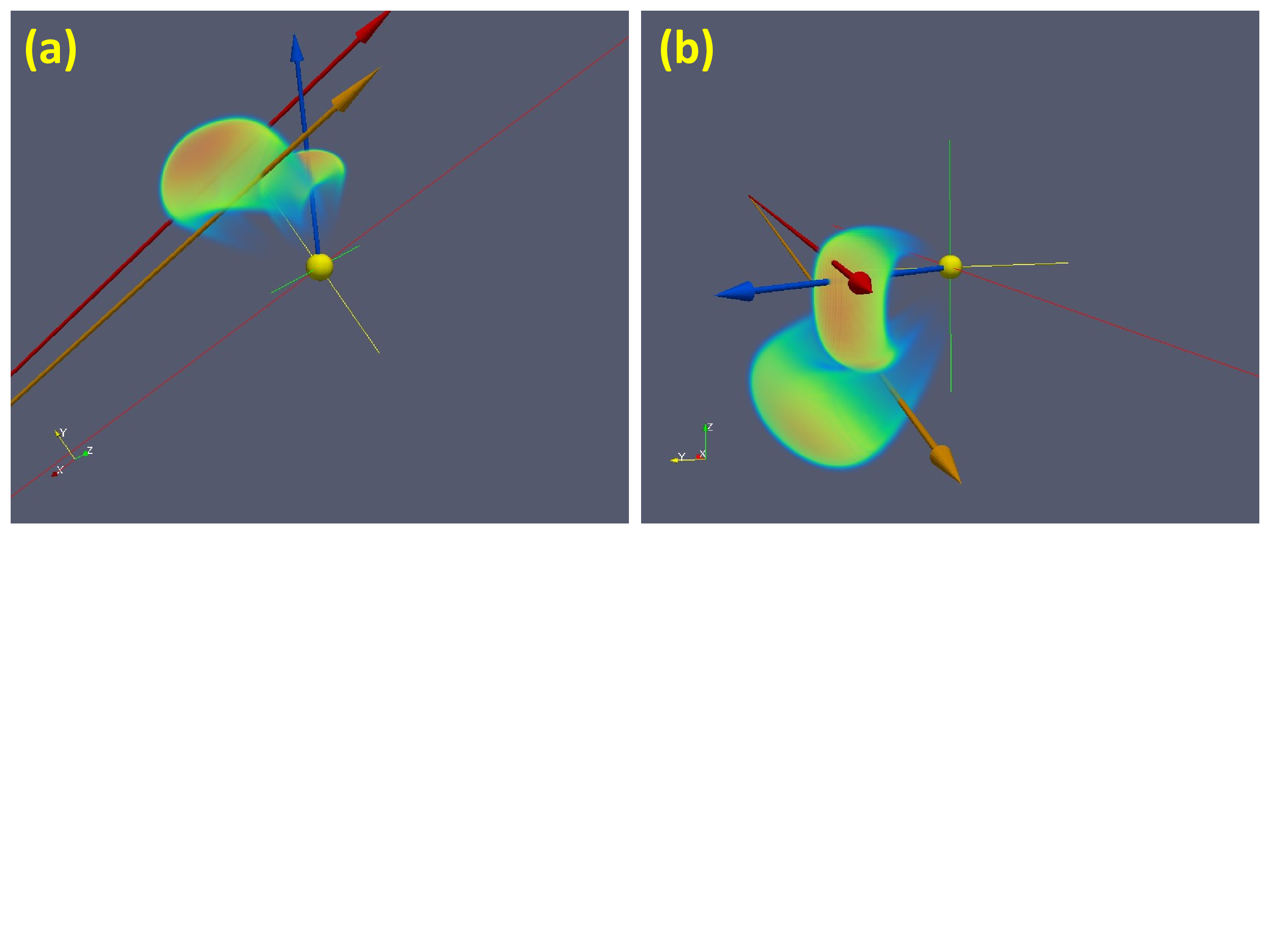}{2.2in}{0}{60}{60}{-213}{-155}
\caption{(a) Reconstructed 3-D MFR structures of CME-1 and CME-2, at
  UT 17:24, shown in HEE coordinates.  CME-1 is the larger, E-W oriented MFR,
  and CME-2 is the smaller N-S oriented MFR.  The
  Sun is at the origin, and is shown to scale.  The blue arrow
  indicates the direction toward {\em STEREO-A}, which goes through CME-2.
  The red and orange arrows indicate the LOS from Earth toward two
  background radio sources, 0842+1835 (red) and 0843+1547 (orange),
  observed by VLA for Faraday rotation purposes.  The 0842 LOS goes through
  CME-2 and the 0843 LOS samples CME-1.  (b) Same as (a), but for a later
  time, UT 19:24.}
\end{figure}
     Simple trial-and-error and subjective judgment are used to vary
the MFR parameters and decide which parameters yield synthetic images
that collectively best match the actual images.  Our final inferred
MFR morphologies of CME-1 and CME-2 are depicted in Figure~4, which
shows the relative positions of the two CMEs at two particular times.
Synthetic images of this morphological reconstruction are shown in
Figures~1 and 2, for comparison with the actual images.  (Movie
versions of these figures in the online article provide a more
comprehensive data/model comparison.)

\begin{deluxetable}{clcc}
\tabletypesize{\scriptsize}
\tablecaption{Flux Rope Parameters}
\tablecolumns{4}
\tablewidth{0pt}
\tablehead{
  \colhead{Parameter} & \colhead{Description}&\colhead{CME-1}&\colhead{CME-2}}
\startdata
$\lambda_s$ (deg)& Trajectory longitude             & 102  & 122  \\
$\beta_s$ (deg)  & Trajectory latitude              &$-40$ &$-10$ \\
$\gamma_s$ (deg) & Tilt angle of MFR                &$-15$ &  80  \\
FWHM$_s$ (deg)   & Angular width                    & 75.8 & 77.1  \\
$\Lambda_s$      & Aspect ratio                     & 0.18 & 0.15  \\ 
$\eta_s$         & Ellipticity of MFR cross section & 1.5  & 1.7  \\
$\alpha_s$       & Shape parameter for leading edge & 3.0  & 5.0  \\
\enddata
\end{deluxetable}
     Table~1 lists the MFR fit parameters for CME-1 and CME-2, using
the variable names from \citet{bew17}.  Briefly, $\lambda_s$ and
$\beta_s$ describe the central trajectories in HEE coordinates, with
the $\lambda_s$ directions explicitly indicated in Figure~3(a).  The
$\gamma_s$ parameter indicates the tilt angle of the MFR, with
$\gamma_s=0^{\circ}$ corresponding to an E-W orientation parallel to
the ecliptic, and $\gamma_s>0^{\circ}$ indicating an upward tilt of
the west leg.  With $\gamma_s=80^{\circ}$, CME-2 is oriented close
to N-S.  This orientation is strongly implied by the clear N-S
asymmetry of the CME halo as viewed in the last COR2-A image in
Figure~2, and the apparent face-on MFR appearance of CME-2 in the COR2-B
and LASCO/C3 images.  Getting this orientation correct is crucial for
interpretation of the {\em STEREO-A} in~situ observations of CME-2 in the
next section.  The FWHM$_s$ parameter is the
full-width-at-half-maximum angular width of the MFR.  The aspect
ratio, $\Lambda_s$, indicates the minor radius of the apex of the MFR
divided by the distance of the apex from the Sun, and so is a measure
of how fat the MFR is.  The ellipticity of the MFR channel is
described by $\eta_s$, which is the major radius divided by the minor
radius.  A value of $\eta_s=1$ would indicate a circular MFR cross
section.  Finally, the $\alpha_s$ parameter defines the shape of the
MFR leading edge, with higher values leading to flatter leading edges.

\section{Reconstruction of CME Field Structure from In~situ Data}

\begin{figure}[t]
\plotfiddle{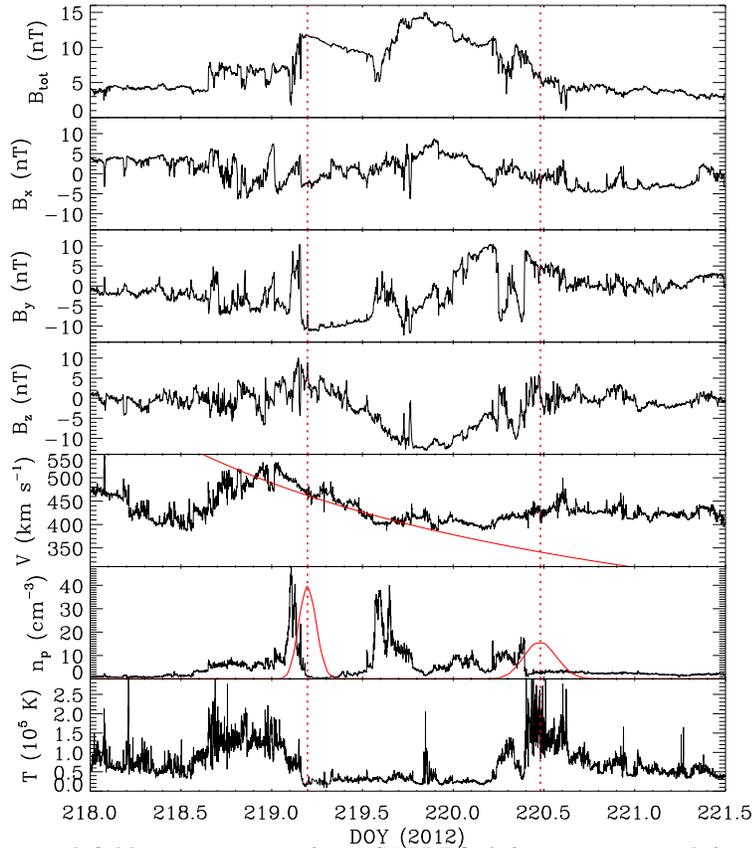}{3.9in}{90}{70}{70}{270}{-60}
\caption{In~situ plasma and field measurements from {\em STEREO-A} for a time
  period from 2012~August~5 (DOY=218) to 2012~August~8 (DOY=221).  The
  top panel shows the total magnetic field, with the individual field
  components shown below it in a spacecraft-centered RTN coordinate
  system.  The bottom three panels are the proton flow speed, number
  density, and temperature.  The solid red lines in the speed and density
  panels are the values predicted from the CME-2 kinematic and
  morphological reconstruction in Figures~3-4.  The vertical red dotted
  lines indicate the predicted MFR encounter time, based on the location
  of the predicted density peaks.}
\end{figure}
     With the morphological and kinematic reconstruction of the
2012~August~2 CMEs complete, we now turn our attention to the
in~situ observations of CME-2 from {\em STEREO-A}, which we use to model the
magnetic field structure of the CME.  The {\em STEREO} in~situ observations
are made by two separate instruments: the Plasma and Suprathermal Ion
Composition (PLASTIC) instrument \citep{abg08}, and the
In~situ Measurements of Particles and CME Transients (IMPACT)
package \citep{mha08,jgl08}.  Figure~5
displays the {\em STEREO-A} in~situ observations for a 3.5 day period
starting on 2012~August~5, corresponding to day-of-year (DOY) 218.
The top four panels show the magnetic field measurements, both the
total field strength, $B_{tot}$, and the individual $B_x$, $B_y$, and
$B_z$ field components in a spacecraft-centered RTN coordinate
system.  The bottom three panels show the proton speed, number
density, and temperature.

     The vertical dotted lines indicate the predicted encounter
time of the CME-2 MFR based on the morphological and kinematic
analysis shown in Figures~3 and 4.  This corresponds nicely to a
period of enhanced $B_{tot}$ and decreased $T$, which are often
interpreted as ICME signatures, leading us to conclude that this is
indeed the in~situ signature of CME-2.  The velocity and density
profiles predicted by the MFR reconstruction are explicitly shown in
the velocity and density panels of Figure~5.  The velocity decrease
predicted by the MFR reconstruction is a consequence of the assumed
self-similar expansion of the MFR, which leads to trailing parts of
the MFR expanding more slowly than the leading edge.  The predicted
velocity profile agrees very well with the observed velocities early
in the ICME period, but the observed velocities deviate to higher
values at later times.

     Since our 3-D reconstruction only places mass on the surface of
the MFR, with a Gaussian density profile of narrow width, the
predicted density signature at {\em STEREO-A} is simply two Gaussian peaks
corresponding to the times when the spacecraft enters and then exits
the MFR.  There is a sharp density peak very near the predicted time
of arrival of the ICME at {\em STEREO-A}.  The predicted peak is only about
2.5 hours too late.  This is a surprising degree of success for the
kinematic model in Figure~3(b), considering that it is based only on
coronagraph observations close to the Sun, therefore requiring almost
four days of extrapolation to {\em STEREO-A} in the absence of additional
constraints from {\em STEREO}'s heliospheric imagers.

    Even more surprising is that the ICME encounter time looks like
a plausible match to the data, corresponding roughly to the duration
of enhanced $B_{tot}$, and the duration of low $T$.  We have more
commonly found that MFR shapes inferred from imaging usually lead to
predicted ICME encounter times at 1~au that are significantly longer
than those actually observed \citep{bew17}, meaning that there
is usually a significant mismatch between the size scales of the
inferred MFR structure from imaging and in~situ data.  The mismatch
seems to be less pronounced in this instance.  All of this is
encouraging for our efforts to infer a plausible 3-D field structure
for CME-2 that is consistent with both the imaging and in~situ data.

     The in~situ ICME field structures that are considered to
provide the strongest cases for the MFR paradigm are those classified
as ``magnetic clouds'' (MCs), which possess field rotations consistent
with passage through an MFR with a helical field
\citep{lb81,km86,lfb88,vb98}.
The ICME in Figure~5 does exhibit field rotations, with $B_y$ in
particular changing from negative to positive, but the field
variations are not very smooth, and possess many irregularities.  The
$B_{tot}$ profile is irregular, and there is also a large density peak
in the middle of the ICME region for which we have no explanation.
All of this suggests that the presumed MFR of CME-2 may be disturbed
in a way that could make it difficult to model precisely.  This
might in principle be caused by interactions between CME-1 and CME-2,
which overlap a little in Figure~4.  This modest overlap
develops as CME-2 reaches its final velocity early in the COR2 FOV,
and the degree of overlap is not expected to change much at later
times due to the similar final speeds of the two CMEs.

     Nevertheless, we can still make some strong qualitative
inferences about the field geometry of CME-2 from the field
characteristics seen in Figure~5 \citep[e.g.,][]{tnc19}.
The images of CME-2 strongly imply
a N-S oriented MFR, as shown in Figure~4.  This means that it is the
$B_z$ panel of Figure~5 that should tell us the direction of the MFR's
central axial field, and the data in Figure~5 provide a very clear
answer.  The $B_z$ field in the ICME is negative, meaning that
the axial field of the MFR is to the south.  The $B_y$ field profile
then tells us the handedness of the azimuthal field around the central
axial field.  With $B_y$ being initially negative, but changing to
generally positive halfway through the ICME period, the clear
inference is that the azimuthal field is right-handed.

     This inferred MFR field geometry means that the field at the
outer surface of the CME-2 MFR should be pointed back toward Earth in
Figure~4(b).  There is therefore a very clear expectation for the
rotation measure for the VLA LOS when it first encounters
CME-2.  It {\em must} be positive.  Fortunately, this is exactly what
is observed \citep[see Section~4 and][]{jek17}.  If the rotation
measure had instead been negative in sign, reconciling the radio data
with the CME field structure inferred from the imaging and {\em STEREO-A}
in~situ data would be impossible, which would in turn lead to serious
questions about how images and in~situ observations of CMEs are
generally interpreted in the MFR paradigm.  All this illustrates the
value of multiple observational constraints on CME structure, to
provide definitive tests as to the self-consistency of the
observations with any assumed field geometry for the structure.

     Moving from purely qualitative inferences of field geometry to
more quantitative results requires the use of a physical model for
MFR structure.  For this purpose, we use the recently developed
model of TNC18.  Traditional modeling of
ICME field structure has generally assumed that an MFR can be
approximated locally as an infinite cylinder with a circular
cross section, and with a force-free field \citep[e.g.,][]{rpl90}.
\citet{tnc16} presented a modified modeling
approach that relaxes the force-free assumption, and this model
has now been further expanded to consider elliptical MFR channels,
as opposed to purely circular ones (TNC18).
This is an important advance for our purposes, as CME-1 and CME-2
are here both inferred to have elliptical MFR channels,
with $\eta_s=1.5$ and $\eta_s=1.7$, respectively, and such
ellipticity is by no means uncommon \citep{bew17}.

     The MFR model of TNC18 is developed in a coordinate system
where the MFR is pointed along the y-axis and the flux rope channel
is therefore mapped in the xz-plane.  (See Figure~1 of TNC18.)
In this coordinate system, the field can be reduced to two components,
an axial field, $B_y$, along the y-axis, and an azimuthal field, $B_{\phi}$,
directed around the origin in the xz-plane.  The total field
is then simply $B_{tot}=\sqrt{B_y^2+B_{\phi}^2}$.  It is crucial
to note that in the elliptic-cylindrical coordinate system used by TNC18,
the direction of $B_{\phi}$ is not circular about the origin like in a
normal polar coordinate system, but is instead along the elliptical
contours defined by the ellipticity assumed for the MFR shape, which for
our purposes here is based on that inferred from images
(e.g., $\eta_s=1.7$ for CME-2).

     We here quantify the field using two parameters, the
axial field at the MFR center, $B_t$, and the maximum azimuthal field
at the surface of the MFR, $B_p$.  We now briefly describe how $B_t$ and
$B_p$ relate to the mathematical formalism of TNC18, but we refer the
reader to that paper for details.  The model of TNC18 uses
polynomial expansions to express the axial and azimuthal components
of the current density, with polynomial exponents $m$ and $n$,
respectively.  They also define a quantity, $\tau$, that relates the
axial field at the MFR axis to that at the surface; and they use
a parameter $\delta$ to quantify the ellipticity of the MFR, which
relates to our $\eta_s$ parameter as $\delta=1/\eta_s$.  In TNC18, the
field components for any pair of indices $n,m$ are basically defined
by the quantities $B_n^0$ and $C_{nm}$, where if 
$f_c\equiv (n+1)/(\delta^2+m+1)$, then $B_t=\delta \tau B_n^0$ and
$B_p=-\delta f_c B_n^0/C_{nm}$.  Following the example
case explored by TNC18, we here consider only the pair of indices
$m=0$ and $n=1$, and we assume $\tau=1$, which corresponds to
the case where the axial field simply falls to zero at the MFR surface.

\begin{figure}[t]
\plotfiddle{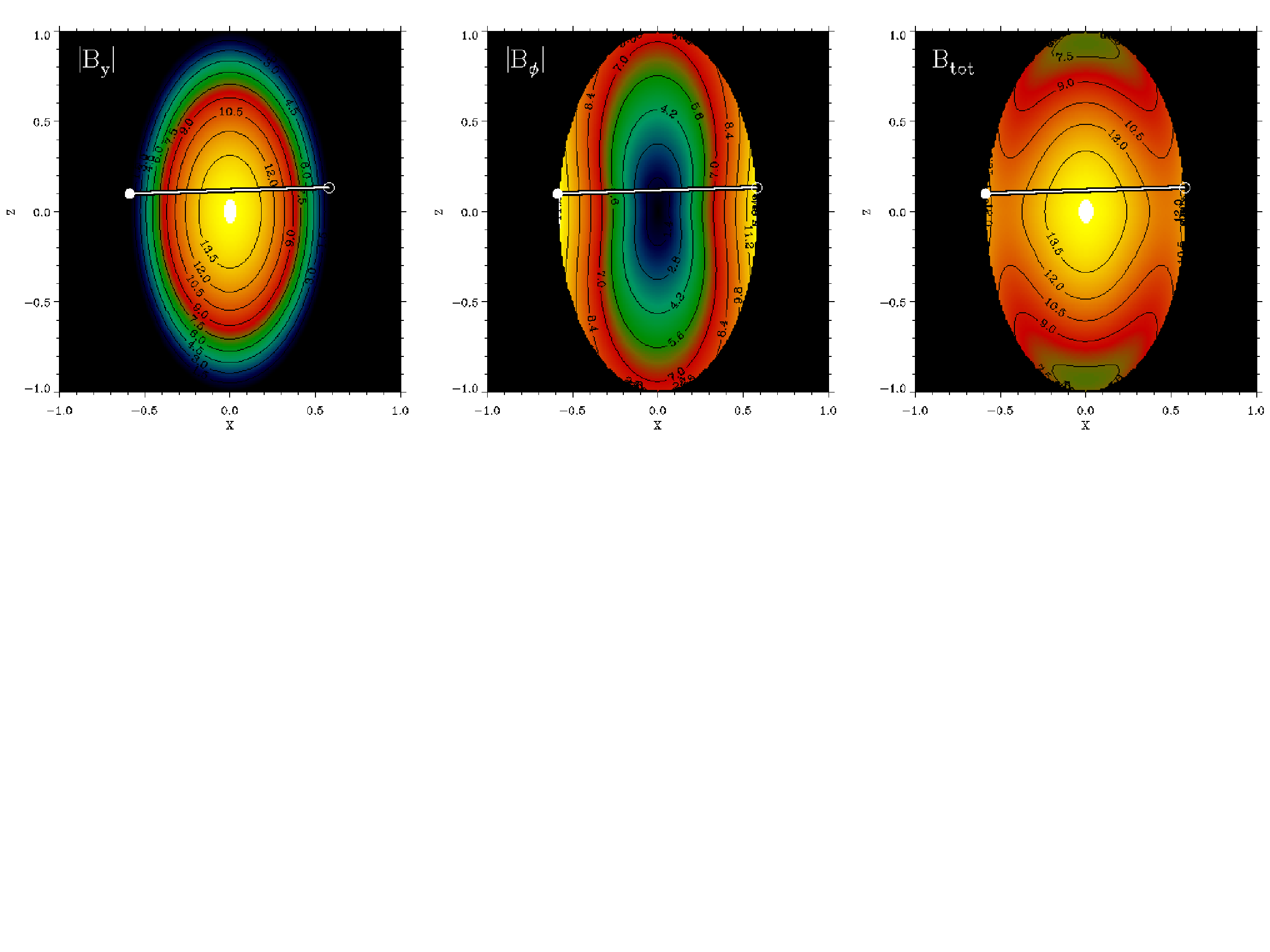}{1.8in}{0}{67}{67}{-242}{-210}
\caption{Maps of the magnetic field in the MFR channel for the model
  that best fits the in~situ data from {\em STEREO-A} (see Figure~7).  From
  left to right, the maps show the axial field, the azimuthal field
  along the elliptical contours of the MFR channel, and the total
  field.  The contours indicate the field values in units of nT.  The
  values shown correspond to the field at the apex of the MFR at the
  time the spacecraft first enters the MFR.  The white line indicates
  {\em STEREO-A}'s path through the MFR channel, entering from the right and
  exiting to the left.}
\end{figure}
     With these assumptions, the TNC18 MFR model for assumed values of
$B_t=+15.0$~nT and $B_p=+12.6$~nT is shown in Figure~6, which we will
show below is a solution that relates to the best fit to the in~situ
field measurements of CME-2 from {\em STEREO-A}.  We avoid ambiguities of
sign for now by simply showing in the figure absolute values of
$B_y$ and $B_{\phi}$.
The axial field, $B_y$, has the peak value of $B_t$ at MFR center, which
decreases to zero at the MFR surface, as required for the $\tau=1$
assumption.  In contrast, $B_{\phi}$ is zero at the center but increases
toward the MFR surface, where we again emphasize that the direction of
$B_{\phi}$ is along elliptical contours about the origin rather than
circular ones.  Its maximum, $B_p$, is at the surface on the minor axis
of the ellipse.  The higher azimuthal fields along the minor axis
compared to the major axis are required by conservation of magnetic flux
about the central axis.

     Our 3-D reconstructed MFR shapes in Figure~4 have cross-sections
that vary along the MFR.  In particular, the cross-sectional area is
large near the apex and smaller in the legs.  A
spacecraft moving through the MFR can in principle pass through
parts of the MFR with different cross-sectional areas, which
we must take into account in modeling the field components that
the spacecraft sees.  The TNC18 MFR model does not
by itself tell us how to do this.  However, a 3-D MFR model must
conserve magnetic flux, both axially and azimuthally, and we use
these flux conservation properties to guide us in relating fields
in one part of the MFR to those in another part.

     We first need to define a reference cross section for the MFR,
and we naturally choose the apex.  At the apex, we use $a_{min}^{ap}$ to
represent the minor radius of the MFR, with field components
$B_y^{ap}$ and $B_{\phi}^{ap}$.  The question then is, what are the
field components at a different part of the MFR with a different
(presumably lower) minor radius, $a_{min}$?  For the axial field, the
integrated magnetic flux will be proportional to the cross-sectional
area of the MFR, which will be proportional to $a_{min}^2$, so at an
arbitrary part of the MFR, $B_y=B_y^{ap}(a_{min}^{ap}/a_{min})^2$.
For the azimuthal field, the integrated magnetic flux for a distance
element $ds$ along the MFR is proportional to $a_{min}B_{\phi}\cdot
ds$.  Assuming the same azimuthal flux for each distance element along
the MFR therefore requires
$B_{\phi}=B_{\phi}^{ap}(a_{min}^{ap}/a_{min})$.  With this assumption,
note that both $B_y$ and $B_{\phi}$ increase as you move down the legs
of the MFR, where the cross-sectional area is lower, but $B_y\propto
1/a_{min}^2$ increases more than $B_{\phi}\propto 1/a_{min}$, meaning
that the overall MFR field becomes more axial in the legs than near
the apex.

     Another complication is that the MFR is not static as the
spacecraft moves through it.  It is after all the CME's radial motion
away from the Sun that is mostly moving the spacecraft through the MFR
structure, and not the spacecraft's heliocentric motion.  The MFR will
be expanding during the spacecraft encounter, and this expansion of
the MFR must also be taken into account.  We have simply assumed
self-similar expansion in the 3-D reconstruction process described in
Section~3.  The defining characteristic of self-similar expansion is
that all distances scale the same in order to preserve the shape of
the expanding structure.  The expansion as a function of time is
described by the kinematic model shown in Figure~3(b), which provides
the leading edge distance of the CME, $R_{le}$, as a useful reference
distance.  Both $a_{min}$ and the overall length of the MFR, $L$, will
scale linearly with $R_{le}$.  Magnetic flux conservation requires
$B_y\propto 1/a_{min}^2$ and $B_{\phi}\propto 1/(a_{min}L)$, so both
$B_y$ and $B_{\phi}$ will be proportional to $1/R_{le}^2$.  Both
components of the magnetic field will therefore decrease with time
in the same manner as the MFR expands.

     The preceding two paragraphs describe how we treat the spatial
and time dependence of the field within the 3-D MFR shapes inferred
from the imaging.  With this established, it is now possible for us
to take a cross-sectional field map like that in Figure~6 for the apex
of the MFR at one time, and extrapolate from it the field structure
of the full 3-D MFR not only at that reference time, but to any other
time as well based on the kinematic model from Figure~3(b).  We can
now proceed to fit the in~situ field observations of CME-2 from
{\em STEREO-A} to find a best-fit field structure in the context of the
TNC18 model.

     A first step is simply to identify the track of the spacecraft
through the MFR, which in Figure~4 is represented as a blue arrow.
White lines in Figure~6 show the {\em STEREO-A} track through the MFR
channel, which moves from right to left.  This is a near direct hit
on the spacecraft, with {\em STEREO-A} passing very close to the axis of
the MFR.  There is a slight positive slope to the track, which is
due to the expansion of the MFR channel during {\em STEREO-A}'s passage
through it, which in a relative sense moves the spacecraft closer to
the MFR's minor axis.

\begin{figure}[t]
\plotfiddle{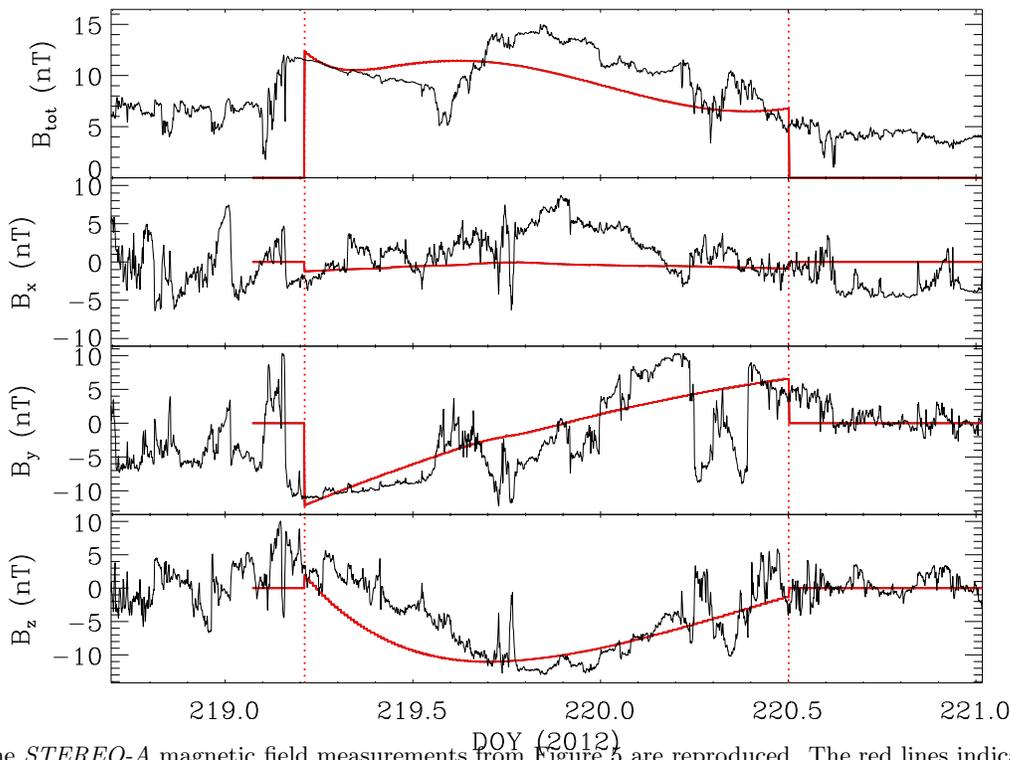}{3.5in}{90}{60}{60}{220}{-50}
\caption{The {\em STEREO-A} magnetic field measurements from Figure~5 are
  reproduced.  The red lines indicate a best-fit to the field values.  The two
  free parameters of the fit are the central axial field ($B_t=+15.0$~nT)
  and the maximum azimuthal field at the MFR surface ($B_p=+12.6$~nT).}
\end{figure}
     Figure~7 reproduces the the {\em STEREO-A} magnetic field measurements
shown in Figure~5.  The goal is to fit these data using the TNC18 MFR
model, operating in the MFR geometry inferred from the image analysis,
with $B_t$ and $B_p$ being the two free parameters.
In doing this, we have to make the coordinate transformations from the
TNC18 coordinate system used in Figure~6 to {\em STEREO-A}'s RTN coordinate
system, which defines the $B_x$, $B_y$, and $B_z$ field components in
Figures~5 and 7.  The best fit to the data is found using $\chi^2$
minimization \citep{prb92}.  We not only fit $B_x$,
$B_y$, and $B_z$ simultaneously, but also $B_{tot}$.  Including
$B_{tot}$ seems redundant, but we see value in considering the
positive-definite $B_{tot}$ quantity in the fit.  Uncertainties for
the $B_{tot}$, $B_x$, $B_y$, and $B_z$ data points must be estimated
to compute the $\chi^2$ quality-of-fit factor.  For $B_{tot}$, we
divide the mean $B_{tot}$ within the MFR encounter time by 30 and take
that as our uncertainty estimate.  For the individual components, we
compute the mean absolute value of $B_x$, $B_y$, and $B_z$ within the
encounter time and divide that by 10 to represent the uncertainty
estimate.  The field variations within the MFR channel are uneven, and
exhibit anomalies that the MFR model cannot hope to fit, such as the
big double-dip decrease in $B_y$ near the end of the MFR encounter,
but we here have made no effort to remove such anomalies before
fitting the data.

     There are only two free parameters of the fit, $B_t$ and $B_p$.
The best fit to the data is shown in Figure~7, with values of
$B_t=+15.0$~nT and $B_p=+12.6$~nT.  As emphasized above, these
values have to be defined for a specific place and time within the
MFR.  The quoted values are for the apex of the MFR at the time when
the MFR first reaches {\em STEREO-A}.  The field maps in Figure~6 correspond
to this particular best-fit field model.  The field values at the actual
location where the spacecraft hits the MFR will be somewhat different,
although in this particular case the difference is tiny.  The general
decrease in model $B_{tot}$ in Figure~7 is due to the expansion of the
MFR during the spacecraft encounter time, which decreases the field
with time.  As a sign convention, we take the polarity of the western
leg of the MFR to be the sign of $B_t$.  For CME-2, which is oriented
close to N-S, the northern leg is slightly west of the southern one,
so a positive $B_t=+15.0$~nT result means that the northern leg is
positive and the southern one negative, leading to the negative $B_z$
seen in the in~situ data.  For $B_p$, positive (negative) values
indicate a right-handed (left-handed) azimuthal field.  Thus, our
$B_p=+12.6$~nT result indicates a right-handed orientation for the
azimuthal field, leading to the rotation of $B_y$ from negative to
positive values in Figure~7.

     In analyses of this nature, we imagine that it could often be
necessary to introduce a couple extra free parameters into the fit
to make the MFR arrival and encounter times at the spacecraft agree
better with the in~situ data.  However, as described above, for this
particular case our MFR reconstruction has ended up with a predicted
arrival time and event duration that already look very plausible, so
no additional scaling parameters are deemed necessary.  We
consider the quality of the fit to the data in Figure~7 to be
impressive considering that there are only two free parameters.
In most past published fits of this nature, there are far more free
parameters \citep[e.g.,][]{rpl11,rpl15}, because there are many
parameters associated with the geometry of the MFR and the
spacecraft's path through it.  In our analysis, the geometry and path
are entirely fixed by the image-based 3-D MFR reconstruction, leaving
only the field parameters to vary.

\section{Comparison with Radio Faraday Rotation Observations}

     We now turn our attention to modeling the radio Faraday rotation
observations for the two lines of sight occulted by the
2012~August~2
CMEs.  The VLA observations are described in detail by \citet{jek17}.
The two background polarized radio sources are both radio
galaxies: Source 0842 (full name 0842+1835) occulted by CME-2, and
source 0843 (full name 0843+1547) occulted by
CME-1.  These lines of sight are shown as red and orange
lines in Figure~4, respectively.  Both lines of sight pass within
about 10 R$_{\odot}$ from Sun-center at closest approach, within the
LASCO/C3 FOV.  \citet{jek17} interpreted the LASCO/C3 images to
suggest that the 0843 LOS grazes the southern edge of CME-2, and is
thereby occulted by both CMEs, although CME-1 clearly dominates the
RM signal.  Our reconstruction here has CME-2 narrowly missing the
0843 LOS, so we will here be intepreting the 0843 data in the context
of CME-1 occultation only.

     The Faraday rotation diagnostic relies on detecting the change
in polarization position angle ($\chi$; defined by the electric field
vector) induced by the passage of a CME in front of the background
source.  This rotation is
\begin{equation}
\Delta\chi=\left[ \left(\frac{e^3}{2\pi m_e^2 c^4} \right)
  \int_{LOS} n_e {\bf B\cdot ds} \right]\lambda^2 = [RM]\lambda^2,
\end{equation}
where $\lambda$ is the observed radio wavelength, and ${\bf ds}$
is the differential direction vector along the LOS.  The term
in square brackets is the rotation measure (RM), which represents
the quantity of interest, with
units of rad~m$^{-2}$.  The constant within the parentheses
includes the electron charge ($e$), the electron mass ($m_e$), and
the speed of light ($c$).
The rotation measure is the integral of the parallel
component of the field times the electron density along the LOS,
multiplied by the constant in parentheses in Equation~(2).
Thus, modeling RM requires both a field model for the CME and
assumptions about the density distribution within it.

\subsection{Analysis of Source 0842 and CME-2}

     We first discuss the source 0842 observations, which provide
RM diagnostics for CME-2.  The {\em STEREO-A} observations of CME-2
have allowed us to model the field of this CME based on the in~situ
data alone, in the context of the TNC18 MFR model and the CME-2 MFR
shape inferred from the images (see Section~3).  Thus, with the
CME field structure already fixed, the only degrees of freedom
we have for modeling RM for source 0842 are those associated with
the CME density.

\begin{figure}[t]
\plotfiddle{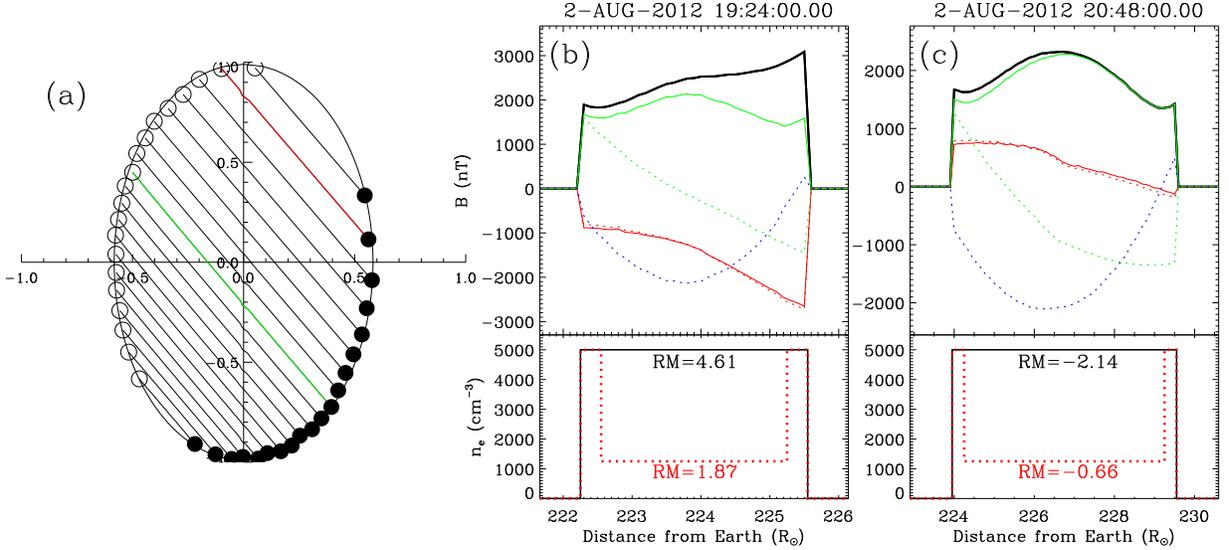}{2.8in}{90}{68}{68}{270}{-100}
\caption{(a) Traces of the LOS to source 0842 through the model MFR
  channel of CME-2 as a function of time, for 12-minute increments
  starting at UT 19:12 on 2012~August~2, with the LOS from Earth
  entering from the top left (open circles) and exiting toward the
  bottom right (filled circles),
  based on the 3-D
  reconstruction in Figure~4.  (b) The red, green, and blue dotted
  lines are traces of $B_x$, $B_y$, and $B_z$, respectively, in GSE
  coordinates for the red track through the MFR channel in (a) based
  on the MFR reconstruction inferred by fitting {\em STEREO-A} in~situ data.
  The red and green solid lines are the projections along the LOS,
  $B_{\parallel}$ and $B_{\perp}$, respectively.  The black line is
  the total field, $B_{tot}$.  The lower panel
  shows two assumed electron density profiles, and the RM observations
  that result from them (in rad~m$^{-2}$ units).  (c) Analogous to
  (b), but for the green track in panel (a).}
\end{figure}
     The first step in modeling RM is to determine the track of
the observed LOS through the CME as a function of time.  We
can do this thanks to the full 3-D reconstruction of the MFR outline
provided by the morphological analysis described in Section~2.  Results are
shown in Figure~8(a).  The figure shows schematically the path of the
LOS to source 0842 through the MFR channel as a function of time, in
12-minute increments.  The LOS from Earth enters the MFR on the top left
and and then traverses the MFR channel toward the bottom right.  The
paths move downwards relative to the MFR as a function of time, passing
through the center of the MFR at about UT 20:24.  The LOS
trajectory is also illustrated in Figure~4(b), which shows the LOS
entering the MFR on the side of the channel and exiting near the
leading edge of the CME, consistent with the early paths in Figure~8(a).

     With the tracks through the MFR established, we can then use
the time-dependent, 3-D MFR model described in Section~3 to trace the
field components along the LOS as a function of time.  This is shown in
Figure~8(b-c) for two example time steps.  The dotted lines in the
upper panels indicate $B_x$, $B_y$, and $B_z$.  These components are
provided in a geocentric-solar-ecliptic (GSE) coordinate system, with
Earth at the origin, the x-axis pointed at the Sun, the z-axis toward
ecliptic north, and the y-axis to the right as viewed from Earth
to form an orthogonal system.  This seems a natural coordinate system
to use given the Earth-based location of the VLA observations.

     For RM computation purposes, we need to know the field direction
relative to the LOS, so Figure~8(b-c) also shows the
$B_{\parallel}$ and $B_{\perp}$ components of the field relative
to the LOS, with $B_{\parallel}$ obviously being the quantity of
particular interest.  Neither $B_{\parallel}$ nor $B_{\perp}$
are really signed quantities, but in the figure we give $B_{\parallel}$
a sign consistent with $B_x$ to emphasize that the two are
nearly identical in magnitude.  This is because the LOS to the
background sources
that VLA is monitoring are naturally very close the Sun, and therefore
they are pointed roughly along the GSE x-axis toward the Sun.

     The second quantity needed to compute model RM values, after
$B_{\parallel}$, is the electron density, $n_e$.  Unlike $B_{\parallel}$,
which is entirely fixed by the Section~3 analysis, $n_e$ is unconstrained.
In the bottom panels of Figure~8(b-c) there
are two different assumed density models.  Both density
models assume a maximum value of $n_e=5000$~cm$^{-3}$.  This value
is roughly consistent with a $1/r^2$ extrapolation of the $n_p$ densities
seen by {\em STEREO-A} at 1~au back to the $\sim 12$~R$_{\odot}$ distance
where the 0842 LOS is encountering the CME.  One
density model (the black line) simply assumes this as a constant
density throughout the MFR.  The other density model, which we call
the shell model (the red dotted line), assumes the peak density is only
at the surface of the MFR, with densities a factor of 4 lower in the
interior.  The RM values computed for the two sample LOS tracks are
provided explicitly in the lower panels of Figure~8(b-c) for both
density models.  In computing these values, it is crucial to note that
in radio astronomy the convention is for RM to be positive for fields
directed along the LOS {\em toward} Earth.  This is opposite from the
$B_{\parallel}$ (and $B_x$) sign in the upper panels of Figure~8(b-c),
which is associated with the GSE coordinate system instead of the
radio convention.  Thus, the sign of RM is reversed from that suggested
by $B_{\parallel}$ in the figure.

\begin{figure}[t]
\plotfiddle{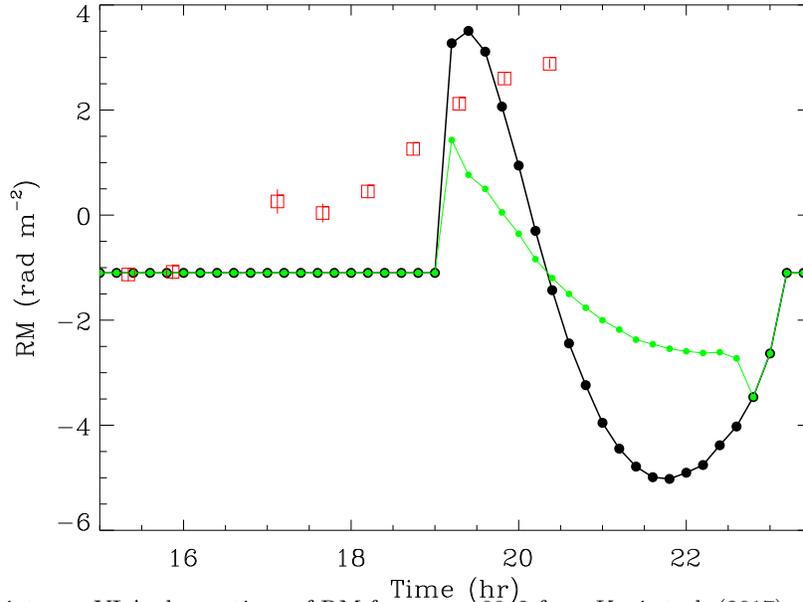}{2.8in}{90}{50}{50}{200}{-45}
\caption{The red data points are VLA observations of RM for source
  0842 from Kooi et al.\ (2017).  The black and green lines are
  predicted RM values using the reconstructed MFR structure based on
  the imaging and {\em STEREO-A} in~situ data, and the two electron density
  profiles in Figure~8(b-c), with the black line assuming a constant
  density in the MFR channel, and the green line assuming density
  peaks at the surface.}
\end{figure}
     Figure~9 shows the model RM values as a function of time for
both the constant density model and the shell model, and compares
them with the RM observations provided by VLA.  As noted above,
refer to \citet{jek17} for details of the VLA observations.
The VLA observations suggest that even outside the CME encounter
there is a background RM value of about $-1.1$ rad~m$^{-2}$.
This is to be expected, as the quiescent solar wind itself will
produce an RM signal.  Our model RM values are displayed after
adding to it this $-1.1$ rad~m$^{-2}$ background value.

     Our density and field MFR parameters can be compared with
those estimated by \citet{jek17}.  Our $n_e=5000$~cm$^{-3}$
value is close to the value of $n_e=6900\pm 500$~cm$^{-3}$ estimated by
\citet{jek17}.  Comparing the field values is somewhat
trickier.  Our $B_t=+15.0$~nT and $B_p=+12.6$~nT values correspond
to a time when the MFR edge reaches {\em STEREO-A}, at which
time the center of the MFR is at about 0.8~au.  Extrapolating
back to a time when the MFR center is at the $\sim 12$~R$_{\odot}$
encounter distance of the 0842 LOS yields values of
$B_t=+31$~mG and $B_p=+26$~mG, which are roughly compatible with
the $B_{CME}=10.4\pm 0.4$~mG value from \citet{jek17},
albeit somewhat higher.  (For reference, note that
1 nT equals 0.01 mG.)

     In comparing the model and observed RM values for CME-2 in
Figure~9, we first emphasize the principle success of the model,
namely that the model RM values are positive, consistent with
the VLA observations.  If the observed RM values had been negative,
there would have been no way whatsoever to reconcile the MFR model
constructed from our interpretation of the images and {\em STEREO-A}
in~situ data with the VLA data.  This would have called into serious
question the standard methodologies by which we and others interpret
CME images and 1~au in~situ data.  The successful prediction of
positive RM values is therefore of significant importance.  The
model prediction of positive RM values by VLA is truly a falsifiable
prediction for the VLA data, and the model passes this crucial test.

     However, there are two worrisome inconsistencies between the
model and observed RM values.  One is that the observed RM values
actually increase about two hours before the MFR reconstruction
suggests that CME-2 occults the LOS.  Inspection of Figure~8 from
\citet{jek17} shows that the LASCO/C3 brightness values at
the source 0842 location do not really start increasing until
after UT 18:30, so it is not surprising that our image-based MFR
reconstruction does not predict any RM signal until after this
time.  We here interpret the premature RM increase as being due
to a sheath region of deflected solar wind field out ahead of the
CME-2 MFR.

     The {\em STEREO-A} in~situ data in Figure~5 actually suggest the
existence of just such a precursor sheath region, between DOY=218.6
and DOY=219.0.  This time range is ahead of the actual time period
of the ICME encounter, which presumably begins with the big
density peak at DOY=219.1, where the low $T$ and high $B_{tot}$
values truly start.  There is clearly a region of disturbed
$B$ and slightly enhanced $n_p$ ahead of this time.  For a relatively
slow CME like CME-2, there is not really any shock associated with
the CME, but there will still be a region of disturbed solar wind
and deflected $B$ out ahead of the CME, which could affect the RM
values observed by VLA even before the CME actually occults the LOS.
[See, e.g., \citet{ek17} for a discussion of ICME shock sheaths
and sheath-like regions.]
This is therefore our interpretation for why the observed RM
values in Figure~9 increase relative to the RM=$-1.1$ rad~m$^{-2}$
background even before the CME arrival.  However, if a sheath region
is affecting the RM values, this is potentially a significant source
of uncertainty regarding the interpretation of the RM values that
are observed even {\em after} the CME actually occults the LOS, given
that the sheath of deflected ambient solar wind field could be
affecting those values as well.

     The second worrisome inconsistency between the model and observed
RM values in Figure~9 is that the model clearly requires that the
sign of RM change from positive to negative relatively quickly.
Note that both the constant density and shell model RM predictions
suggest a change in RM sign at about UT 20:20, roughly when the LOS
passes through the center of the MFR channel (see Figure~8(a)).
No change in the assumed density model will affect this prediction
significantly, as it is determined almost entirely by the field model,
which is constrained by the {\em STEREO-A} in~situ data.  In contrast, the
actual VLA observations do not show any indication that the positive
RM values reverse sign.  Unfortunately, the VLA data end too early,
the last VLA data point being at UT 20:20.  This makes it very hard
for us to assess the seriousness of the discrepancy.  Even just one
more hour of VLA data would have been very illuminating.  If that hour
had shown the beginnings of an RM sign reversal, we could have
concluded that this is consistent with the model and that the delayed
reversal is simply due to inaccuracies in the field model resulting
from the fit to the in~situ data in Figure~7.  However, if that extra
hour had shown no evidence for any reversal whatsoever, we would have
concluded that this is a serious inconsistency with the single MFR
reconstruction of CME-2, and that our interpretation of the images
and {\em STEREO-A} in~situ data for this event is therefore in serious
doubt as a consequence.  In the absence of this extra hour of VLA
data, we instead have to consider the issue as still being ambiguous.

     Finally, it should be noted that the shell model for the CME
densities is clearly not favored by the data in Figure~9.  The shell
model predicts rather sharp maxima in density when the CME first occults
the LOS, which are not observed.  This is also the case for CME-1,
although we will not demonstrate that explicitly in the next section.
For CME-1 we will exclusively assume a constant density model.
This provides evidence that the shell model is a worse
approximation for these CMEs than the constant density model.

\subsection{Analysis of Source 0843 and CME-1}

\begin{figure}[t]
\plotfiddle{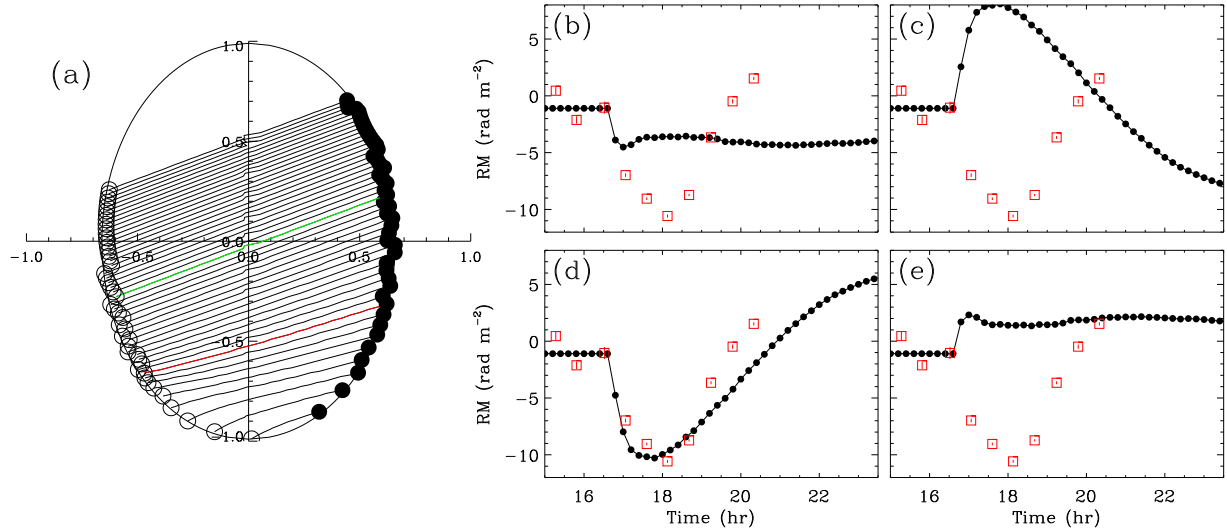}{2.8in}{90}{68}{68}{270}{-100}
\caption{(a) Traces of the LOS to source 0843 through the model MFR
  channel of CME-1 as a function of time, for 12-minute increments
  starting at UT 16:48 on 2012~August~2, with the LOS from Earth
  entering from the left and exiting to the right, based on the 3-D
  reconstruction in Figure~4.  The red and green lines are paths
  referred to in Figure~11.  (b-e) The red data points show the VLA RM
  values of source 0843 from Kooi et al.\ (2017).  The black lines
  are model RM values assuming the following magnetic field component
  values: (b) ($B_t$,$B_p$)=(+10,+10)~nT, (c) ($B_t$,$B_p$)=(+10,-10)~nT,
  (d) ($B_t$,$B_p$)=(-10,+10)~nT, and (e) ($B_t$,$B_p$)=(-10,-10)~nT.
  The data clearly favor the field polarity assumed in (d).}
\end{figure}
     We now turn our attention to the source 0843 data, and CME-1.
Figure~10(a) shows the paths of the LOS through the model CME-1 MFR
channel, in 12-minute increments starting at UT 16:48 on 2012~August~2
and extending for 9.2 hours thereafter.  Figure~4(a) provides another
illustration of how CME-1 is encountering the LOS.  The eastern leg of
the CME is tilted north of the LOS and therefore the VLA observations
of 0843 basically end up probing the western half of the CME, with the
LOS at later times moving down the western leg of the MFR as the CME
expands outwards.  The expansion eventually moves the LOS through the
center of the MFR channel, at about UT 21:24.

     In contrast to the CME-2 analysis, there is no 1~au in~situ data
to constrain the MFR model.  Thus, there are more free parameters
to consider when trying to reproduce the VLA RM observations.
However, the source 0843 RM constraints on CME-1 shown in
Figure~10(b-e) are, by themselves, better than those for CME-2 in two
respects.  One is that the RM values are simply larger, and the other
is that the sign of RM is observed to reverse at about UT 19:30, which
is very helpful for constraining the field model of the MFR even in
the absence of 1~au in~situ constraints.

     In Figure~10(b-e), we show the RM values predicted by
an MFR model assuming four possible polarities for the MFR field; in
particular with ($B_t$,$B_p$)=(+10,+10)~nT,
($B_t$,$B_p$)=(+10,-10)~nT, ($B_t$,$B_p$)=(-10,+10)~nT, and
($B_t$,$B_p$)=(-10,-10)~nT.  Analogous to what we did for the CME-2
analysis, these values are for the apex of the MFR, for the time when
the leading edge of the MFR first reaches 1~au, and we are once again
assuming a background RM value of $-1.1$ rad~m$^{-2}$.  We are using
the 1~au arrival time as the reference time, even in the absence of
any relevant data at that time for this CME, in order to allow for easier
comparison with the ($B_t$,$B_p$)=(+15.0,+12.6)~nT best-fit values
found for CME-2 (see Figure~7).  As for the density model, in
Figure~10(b-e) we simply assume a constant density within the MFR
channel of $n_e=10,000$~cm$^{-3}$, a value about a factor of two
lower than estimated by \citet{jek17}.

     The (-,+) polarity model in Figure~10(d) is by far the favored
polarity.  The (+,-) and (-,-) polarities of panels (c) and (e)
are clearly wrong, given that they yield initial positive
RM values instead of the observed negative values.  The (+,+)
polarity of panel (b) does yield negative RM values, but the
model shows no hint of reversing the RM sign at later times,
in contrast with the data.  Thus, the panel (d) model is clearly
best, with the ($B_t$,$B_p$)=(-10,+10)~nT field model combined
with the constant-density $n_e=10,000$~cm$^{-3}$ model yielding
a plausible match to the data.  It should be emphasized that in the
absence of 1~au constraints, there is a degeneracy between the field
and density values.  If we arbitrarily increased the field by a factor
of 2 and decreased the density by a factor of two, the model RM
predictions would be the same, for example.

\begin{figure}[t]
\plotfiddle{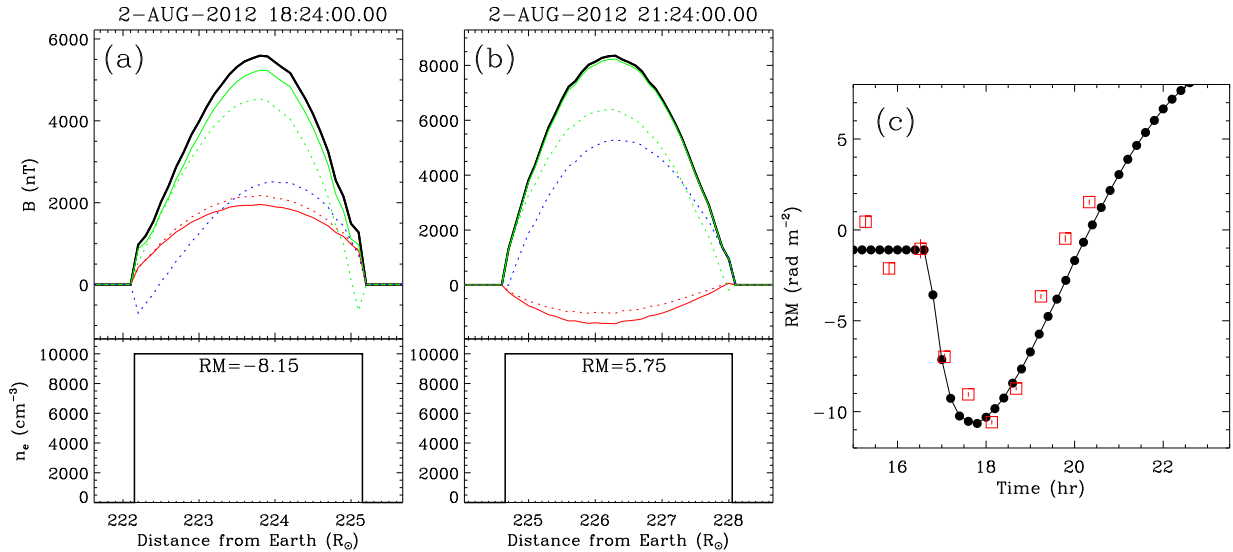}{2.8in}{90}{68}{68}{270}{-100}
\caption{(a) The red, green, and blue dotted lines are traces of
  $B_x$, $B_y$, and $B_z$, respectively, in GSE coordinates for the
  red track through the model CME-1 MFR channel from Figure~10(a),
  assuming a field model with ($B_t$,$B_p$)=(-20,+5)~nT.  The red and
  green solid lines are the projections along the LOS, $B_{\parallel}$
  and $B_{\perp}$, respectively.  The black line is the total field,
  $B_{tot}$.  The lower panel shows the electron density profile
  assumed to compute the displayed RM value (in rad~m$^{-2}$ units).
  (b) Analogous to (a), but for the green track in Figure~10(a).  (c)
  The model RM values for source 0843 occulted by CME-1 as a function
  of time (black line), compared with the VLA observations (red
  boxes).}
\end{figure}
     In Figure~11, we present a model that provides a somewhat better
fit to the observations than Figure~10(d), with the RM sign change
pushed to an earlier time.  This model assumes
($B_t$,$B_p$)=(-20,+5)~nT, with the densities remaining at
$n_e=10,000$~cm$^{-3}$.  Figure~11(a-b) are analogous to
Figure~8(b-c), explicitly showing the field components of
this best-fit model for two example time steps.  The second time
step is at the time the LOS crosses the center of the MFR, by
which time the sign of RM has already reversed
from its initial sign.  Figure~11 shows the fit to the data
resulting from this model, which is a much better fit than we
found for the CME-2 data.  However, the better quality fit is
in part allowed by the freedom we have to assume field model
parameters to be anything we want, in absence of any 1~au in~situ
constraints.  Nevertheless, the model is clearly more
successful in reproducing the arrival time of the CME at the LOS.
There is no precursor RM signature like we found for in the
CME-2 analysis, which we interpreted as indicative of an RM
signature of a sheath region.

     The (-,+) polarity of the CME-1 MFR implies that the
central axial field of the roughly E-W oriented MFR is
directed from east to west, and the azimuthal field about
the axis is right-handed.  The western leg of CME-1 and the
southern leg of CME-2 partly overlap in Figure~4.  These
are both negative polarity legs in our MFR reconstructions.


\section{Summary}

     We have studied the magnetic field structures of two CMEs
(CME-1 and CME-2) that erupt from the west limb as viewed from Earth
on 2012~August~2, for which there are uniquely extensive observational
constraints.  These constraints include excellent stereoscopic imaging
observations from {\em STEREO} and {\em SOHO}/LASCO, allowing reconstruction of the
3-D MFR structure of the two events.  Both CMEs also occult two
separate lines of sight observed by VLA
(sources 0843 and 0842, respectively),
which provide constraints on the field strength and orientation of
the two CMEs.  Finally, CME-2 hits {\em STEREO-A} on
August~6, making this CME particulary notable as being the first CME
with constraints from stereoscopic imaging, 1~au in~situ data, and
radio RM observations.  Our analysis considers all of these data,
in the context of a physical MFR model from TNC18.  Our findings are
summarized as follows:
\begin{description}
\item[1.] For CME-2, the combination of imaging and {\em STEREO-A}
  in~situ data allows a full time-dependent 3-D model of the MFR field
  structure to be constructed.  This best-fit model has a central
  axial field, $B_t$, and maximum surface azimuthal field, $B_p$, of
  ($B_t$,$B_p$)=(+15.0,+12.6)~nT.  These values are quoted for the MFR
  apex at a time when the leading edge of that apex reaches {\em STEREO-A}.  A
  density model simply assuming a constant density within the MFR
  channel, with $n_e=5000$~cm$^{-3}$, seems to work better for reproducing
  the RM measurements than a
  shell model with peak densities at the surface of the CME.
\item[2.] The CME-2 MFR reconstruction makes two clear falsifiable
  predictions for the radio RM observations: 1. The sign of RM
  observed by VLA toward source 0842 must be positive, and 2. The
  sign of RM should reverse to negative within a couple hours.  The
  VLA data pass the first test.  This is significant, as a negative
  result would have invalidated the single MFR model of the CME,
  potentially casting doubt on the kinds of assumptions that are often
  made when interpreting images of CMEs and 1~au in~situ data.
  However, the quality of the overall fit to the observed RM values is not
  great.  In particular, the VLA RM observations do not seem
  to pass the second test, showing no indication of a sign reversal,
  but unfortunately the data end too early to be sure of the significance
  of this discrepancy.
\item[3.] For CME-2, the RM values actually increase a couple hours
  prior to when the MFR reconstruction predicts that the CME
  occults the 0842 LOS.  We interpret this as being due to a sheath
  region of deflected field out ahead of the MFR that is itself
  producing an RM signature.  The existence of RM signatures from
  CME sheaths could significantly complicate interpretations of
  radio Faraday rotation from CMEs.
\item[4.] Although there are no 1~au in~situ constraints for CME-1,
  the radio data by themselves provide strong constraints on
  the field geometry of the CME, since the RM observations show a
  sign reversal, the timing of which has substantial diagnostic
  power.  Our best-fit model has ($B_t$,$B_p$)=(-20,+5)~nT, where
  we here quote the values extrapolated to 1~au to allow for ease
  of comparison with the CME-2 fit.  With a simple assumption
  of constant density throughout the MFR, the best-fit model
  has $n_e=10,000$~cm$^{-3}$.
\item[5.] The model MFR density and field values of CME-1 and CME-2
  reported here are within roughly a factor of $2-3$ of those
  inferred in the previous analysis of \citet{jek17}, an
  acceptable level of agreement considering the different
  underlying assumptions of these independent analyses.
\end{description}

     This study reveals the promise of joint radio, white light
imaging, and in~situ studies of CMEs, although obtaining more such
observations in future observing campaigns will rely on good fortune,
given the unpredictable nature of solar transients.  Further progress
could also be achieved by the monitoring of a larger number of
background radio sources behind a CME, allowing for the development
of a more detailed model of 3-D magnetic field structure, even in the
absence of in~situ field measurements from an encounter with a
spacecraft.

\acknowledgments

Financial support was provided by the Chief of Naval Research, and by
NASA award 80HQTR18T0084 to the Naval Research Laboratory.  The
{\em STEREO}/SECCHI data are produced by a consortium of NRL (US), LMSAL (US),
NASA/GSFC (US), RAL (UK), UBHAM (UK), MPS (Germany), CSL (Belgium),
IOTA (France), and IAS (France).  In addition to funding by NASA, NRL also
received support from the USAF Space Test Program and ONR.
This work has also made use of data provided by the {\em STEREO} PLASTIC and
IMPACT teams, supported by NASA contracts NAS5-00132 and NAS5-00133.


\begin{thebibliography}{}

\bibitem[Acu\~{n}a et al.(2008)]{mha08}
Acu\~{n}a, M. H., Curtis, D., Scheifele, J. L., et al. 2008,
  Space Sci.~Rev., 136, 203
\bibitem[Al-Haddad et al.(2013)]{nah13}
Al-Haddad, N., Nieves-Chinchilla, T., Savani, N. P., et al. 2013,
  Sol.~Phys., 284, 129
\bibitem[Bevington \& Robinson(1992)]{prb92}
Bevington, P. R., \& Robinson, D. K. 1992, Data Reduction and Error
  Analysis for the Physical Sciences (New York: McGraw-Hill)
\bibitem[Billings(1966)]{deb66}
Billings, D. E. 1966. A Guide to the Solar Corona (New York: Academic
  Press)
\bibitem[Bird et al.(1980)]{mkb80}
Bird, M. K., Schruefer, E., Volland, H., \& Sieber, W. 1980, Nature, 283, 459
\bibitem[Bird et al.(1985)]{mkb85}
Bird, M. K., Volland, H., Howard, R. A., et al. 1985, Sol.~Phys., 98, 341
\bibitem[Bisi et al.(2016)]{mmb16}
Bisi, M. M., Jensen, E., Sobey, C., et al. 2016,
  Abstract SH11C-2251 presented at 2016 AGU Fall Meeting
\bibitem[Bothmer \& Schwenn(1998)]{vb98}
Bothmer, V., \& Schwenn, R. 1998, Ann.~Geophys., 16, 1
\bibitem[Brueckner et al.(1995)]{geb95}
Brueckner, G. E., Howard, R. A., Koomen, M. J., et al. 1995, Sol.~Phys.,
  162, 357
\bibitem[Burlaga(1988)]{lfb88}
Burlaga, L. F. 1988, JGR, 93, 7217
\bibitem[Burlaga et al.(1981)]{lb81}
Burlaga, L., Sittler, E., Mariani, F., \& Schwenn, R. 1981, JGR,
  86, 6673
\bibitem[Eyles et al.(2009)]{cje09}
Eyles, C. J., Harrison, R. A., Davis, C. J., et al. 2009, Sol.~Phys., 254, 387
\bibitem[Galvin et al.(2008)]{abg08}
Galvin, A. B., Kistler, L. M., Popecki, M. A., et al. 2008,
  Space Sci.~Rev., 136, 437
\bibitem[Gibson \& Low(1998)]{seg98}
Gibson, S. E., \& Low, B. C. 1998, ApJ, 493, 460
\bibitem[Gopalswamy et al.(2009)]{ng09}
Gopalswamy, N., Yashiro, S., Michalek, G., et al. 2009, EM\&P, 104, 295
\bibitem[Howard et al.(2008)]{rah08}
Howard, R. A., Moses, J. D., Vourlidas, A., et al. 2008, Space Sci. Rev.,
  136, 67
\bibitem[Howard et al.(2016)]{tah16}
Howard, T. A., Stovall, K., Dowell, J., Taylor, G. B., \& White, S. M.
  2016, ApJ, 831, 208
\bibitem[Hu \& Sonnerup(2001)]{qh01}
Hu, Q., \& Sonnerup, B. U. \"{O}. 2001, GeoRL, 28, 467
\bibitem[Illing \& Hundhausen(1985)]{rmei85}
Illing, R. M. E., \& Hundhausen, A. J. 1985, JGR, 90, 275
\bibitem[Ingleby et al.(2007)]{ldi07}
Ingleby, L. D., Spangler, S. R., \& Whiting, C. A. 2007, ApJ, 668, 520
\bibitem[Jensen et al.(2018)]{eaj18}
Jensen, E. A., Heiles, C., Wexler, D., et al. 2018, ApJ, 861, 118
\bibitem[Kilpua et al.(2017)]{ek17}
Kilpua, E., Koskinen, H. E. J., \& Pulkkinen, T. I. 2017, LRSP, 14, 5
\bibitem[Kilpua et al.(2012)]{ekjk12}
Kilpua, E. K. J., Mierla, M., Rodriguez, L., et al. 2012, Sol.~Phys., 279, 477
\bibitem[Kooi et al.(2014)]{jek14}
Kooi, J. E., Fischer, P. D., Buffo, J. J., \& Spangler, S. R. 2014,
  ApJ, 784, 68
\bibitem[Kooi et al.(2017)]{jek17}
Kooi, J. E., Fischer, P. D., Buffo, J. J., \& Spangler, S. R. 2017,
  Sol. Phys., 292, 56
\bibitem[Lepping et al.(1990)]{rpl90}
Lepping, R. P., Jones, J. A., \& Burlaga, L. F. 1990, JGR, 95,
  11957
\bibitem[Lepping et al.(2011)]{rpl11}
Lepping, R. P., Wu, C. -C., Berdichevsky, D. B., \& Szabo, A. 2011,
  Sol.~Phys., 274, 345
\bibitem[Lepping et al.(2015)]{rpl15}
Lepping, R. P., Wu, C. -C., Berdichevsky, D. B., \& Szabo, A. 2015,
  Sol.~Phys., 290, 2265
\bibitem[Levy et al.(1969)]{gsl69}
Levy, G. S., Sato, T., Seidel, B. L., et al. 1969, Science, 166, 596
\bibitem[Lugaz et al.(2009)]{nl09}
Lugaz, N., Vourlidas, A., \& Roussev, I. I. 2009, Ann.~Geophys., 27, 3479
\bibitem[Luhmann et al.(2008)]{jgl08}
Luhmann, J. G., Curtis, D. W., Schroeder, P., et al. 2008,
  Space Sci.~Rev., 136, 117
\bibitem[Marubashi(1986)]{km86}
Marubashi, K. 1986, Adv.~Space~Res., 6, 335
\bibitem[M\"{o}stl et al.(2018)]{cm18}
M\"{o}stl, C., Amerstorfer, T., Palmerio, E., et al. 2018, Space Weather,
  16, 216
\bibitem[Nieves-Chinchilla et al.(2019)]{tnc19}
Nieves-Chinchilla, T., Jian, L. K., Balmaceda, L., et al. 2019,
  Sol.~Phys., 294, 89
\bibitem[Nieves-Chinchilla et al.(2016)]{tnc16}
Nieves-Chinchilla, T., Linton, M. G., Hidalgo, M. A., et al. 2016, ApJ, 823, 27
\bibitem[Nieves-Chinchilla et al.(2018)]{tnc18}
Nieves-Chinchilla, T., Linton, M. G., Hidalgo, M. A., \& Vourlidas, A.
  2018, ApJ, 861, 139 [TNC18]
\bibitem[Ord et al.(2007)]{smo07}
Ord, S. M., Johnston, S., \& Sarkissian, J. 2007, Sol.~Phys., 245, 109
\bibitem[Robbrecht et al.(2009)]{er09}
Robbrecht, E., Berghmans, D., \& Van der Linden, R. A. M. 2009, ApJ, 691, 1222
\bibitem[Sakurai \& Spangler(1994)]{ts94}
Sakurai, T., \& Spangler, S. R. 1994, ApJ, 434, 773
\bibitem[Thernisien et al.(2006)]{afrt06}
Thernisien, A. F. R., Howard, R. A., \& Vourlidas, A. 2006, ApJ, 652, 763
\bibitem[Thernisien et al.(2009)]{at09}
Thernisien, A., Vourlidas, A., \& Howard, R. A. 2009, Sol.~Phys., 256, 111
\bibitem[Vandas \& Romashets(2003)]{mv03}
Vandas, M., \& Romashets, E. P. 2003, A\&A, 398, 801
\bibitem[Vourlidas et al.(2014)]{av14}
Vourlidas, A., Lynch, B. J., Howard, R. A., Li, Y. 2014, Sol.~Phys., 284, 179
\bibitem[Wood et al.(2020)]{bew20}
Wood, B. E., Hess, P., Howard, R. A., Stenborg, G., \& Wang, Y. -M. 2020,
  ApJS, 246, 28
\bibitem[Wood \& Howard(2009)]{bew09}
Wood, B. E., \& Howard, R. A. 2009, ApJ, 702, 901
\bibitem[Wood et al.(2010)]{bew10}
Wood, B. E., Howard, R. A., \& Socker, D. G. 2010, ApJ, 715, 1524
\bibitem[Wood et al.(2017)]{bew17}
Wood, B. E., Wu, C. -C., Lepping, R. P., et al. 2017, ApJS, 229, 29
\bibitem[Yashiro et al.(2004)]{sy04}
Yashiro, S., Gopalswamy, N., Michalek, G., et al. 2004, JGR, 109, A07105

\end{thebibliography}
\end{document}